\documentclass[doublecol]{epl2}
\usepackage{graphicx,amsmath,amsfonts,array,mathabx}
\usepackage{enumitem}
\usepackage{braket}
\def\p{\partial}

\def\be{\begin{equation}}
\def\ee{\end{equation}}

\title{Shapes and singularities in triatic liquid crystal vesicles}
\shorttitle{Title} 

\author{Mark J. Bowick\inst{1,2} \and O. V. Manyuhina\inst{1} \and F. Serafin\inst{1,2}}
\shortauthor{F. Author \etal}

\institute{                    
  \inst{1} Physics Department, Syracuse University - Syracuse, NY 13244, USA \\
  \inst{2} Kavli Institute for Theoretical Physics -
University of California, Santa Barbara, CA 93106, USA
}

\abstract{
Determining the equilibrium configuration and shape of curved two-dimensional films with (generalized) liquid crystalline (LC) order is a difficult infinite dimensional problem of direct relevance to the study of generalized polymersomes, soft matter and the fascinating problem of understanding the origin and formation of shape (morphogenesis). The symmetry of the free energy of the LC film being considered and  the topology of the surface to be  determined often requires that the equilibrium configuration possesses singular structures in the form of topological defects such as disclinations for nematic films. The precise number and type of defect plays a fundamental role in restricting the space of possible equilibrium shapes. Flexible closed vesicles with spherical topology and nematic or smectic order, for example,  inevitably  possess  four elementary strength
$+1/2$ disclination defects positioned at the four vertices of a tetrahedral shell. Here we address the problem of determining the equilibrium shape of flexible vesicles with generalized LC order. The order parameter in these cases is an element of $S^1/Z_p$, for any positive integer $p$. We will focus on the case $p =3$, known as \emph{triatic} LCs. We construct the appropriate order parameter for triatics and find the associated free energy. We then describe the structure of the elementary defects of strength $+1/3$ in flat space. Finally, we prove that sufficiently floppy triatic vesicles with the topology of the 2-sphere equilibrate to octahedral shells with strength $+1/3$ defects at each of the six vertices, independently of scale.}

\begin{document}

\maketitle

\section{Introduction}
  {The design of new materials depends on the ability to transfer the properties of the microscopic constituents to the macroscopic scale. Particularly interesting in this context is the synthesis of building blocks of pre-assigned shape, such as ``super-atoms", nanoparticles, and colloids. 
Self-assembly of these elementary blocks can be seen as a generalisation of atomic chemistry at larger scales, with the important difference that here the landscape of the resulting large-scale materials and their properties (chemical, structural, thermodynamic) is much richer, because they often  possess novel symmetry. Many techniques have been employed in recent years to  obtain supramolecular shapes~\cite{Glotzer:2007}. Here we present a purely physical mechanism for naturally generating polyhedral shells. In particular we show that a sufficiently floppy vesicle or spherical interface with long range orientational order in the constituent three-fold symmetric building block~\cite{Dish:2002},~\cite{Jia:2009} inevitably relaxes to an octahedral shell as it equilibrates (for other examples see~\cite{Kar:2008},~\cite{Lube:1992}). The final shape depends only on the ratio of bending rigidity to liquid crystalline elastic moduli and is scale independent, as opposed to buckling phenomena. This mechanism for shape formation is relatively unexplored.\footnote{See \texttt{http://www.sacannagroup.com/home} for examples of the wide variety of novel structures being created by chemists.}
The conceptual framework described can lead to a wide variety of self-assembling polyhedral building blocks. The design rules for anisotropy-driven self-assembly are only just beginning to be understood~\cite{Glotzer:2007}. We also relate the macroscopic polyhedral symmetry group of the ``super-atom" to the microscopic point-symmetry group of the generalized liquid crystal needed for its formation.}
\section{Triatic Liquid crystals}
The behavior of a Liquid Crystal (LC) results from the interplay between its fluid properties, such as response to shear forces, and the intrinsic anisotropy of its microscopic constituents. The tendency to preserve local orientational order arises from the shape of the liquid crystal molecules and their mutual interactions. The oldest case studied is a nematic LC, which models a fluid of rods. The order parameter, an element of $\mathtt{RP}^1$, can be viewed as a unit vector field $\mathbf{p}(\mathbf{x})$ identified with its image under a rotation by $\pi$: $\mathbf{p}\simeq-\mathbf{p}$. It is then natural to ask if there exist models of LC with higher symmetry, such as three, four, and in general $p-$fold symmetry under local rotation of the order parameter. In what follows we will concentrate on the case $p=3$, which will be called triatic. 
\par It is convenient to adopt a coarse-grained description of the system, and introduce an order parameter, $\mathbf Q$, whose properties under spatial rotations model the anisotropic character of the liquid crystal. We will restrict our analysis to perfectly regular molecules in $d=2$ dimensions. 
The nature of the microscopic interactions between molecules is left unspecified, but it is assumed to produce average alignment of the molecules in macroscopic (yet small compared to the characteristic correlation length) portions of the system. The symmetries of the order parameter must be consistent with the resulting anisotropy within a given region. We therefore identify a given state 
of the system with its images under the action of the cyclic group $C_p$, which means under rotations by multiples of $(2\pi)/p$. 
Hence, from the point of view of symmetry, a liquid crystal represents a fluid in which the full rotation group in $d$ dimensions, $SO(d)$, is locally broken to a discrete subgroup of order $p$. 
An alternative, but equivalent, point of view, is to require that at each point in the system, the group $SO(d)$ acts on $\mathbf Q$ modulo $\mathbb{Z}_p$. In most cases, the order parameter is built from a unit vector $\mathbf{p}$ and from the probability density $\rho(\mathbf x,\mathbf n)$ for it to be found along a particular orientation. The lowest non constant moment of the distribution $\rho$ consistent with the symmetries is taken to be the order parameter, which in general is a tensor.
The appropriate tensor for the triatic will be found to be
\begin{equation}
\mathbf {Q}=\sum_{k=0}^2 R^k\mathbf{p}\otimes R^k\mathbf{p} \otimes R^k\mathbf{p},
\end{equation}
where $R^k$ is the rotation of a reference unit vector $\mathbf{p}$ by $k(2\pi)/3$ Third rank tensors have been used as order parameters to describe tetrahedral LC in 3-dimensional space, see for example \cite{Fel:1995}. For experimental realizations of three-fold symmetric LC, see \cite{Exp:2012}.\\
The dynamics of the system is governed by a free energy $F=\int_\Sigma f(\mathbf Q) d^2\sigma$, functional of the order parameter, that will be constructed according to the framework of Landau theory: it is the most general expression formed from the invariants of the $\mathbf Q$ tensor, and each term must be compatible with the local symmetries of the system. 
\par We want to determine the equilibrium (or ground) state of the system at zero temperature. In particular, we are interested in the structure of the minimally defected ground states allowed by the symmetries. We will do it in two settings: at first, the triatic will be confined to an infinite flat space. Then we will study a closed fluid membrane whose surface supports triatic LC order. The latter system has interesting ground state configurations, resulting from the coupling between the order parameter and the geometry of the substrate, as now the free energy contains interaction terms between $\mathbf Q$ and the metric tensor $g_{\mu\nu}$ on the surface. The structure of the defects plays a crucial role in determining the optimal shape of the membrane. 
\section{Order parameter for triatic}
In this section we will discuss the origin of the order parameter suitable for triatic liquid crystals. We will work in analogy with the nematic $Q-$tensor.
Nematic order in $d=2$ is described by an element of $\mathtt{RP}^2$, identifying antipodal points of the unit circle: $\mathbf{p}\simeq\mathbf{-p}$, $\mathbf{p}\in S^1$. The probability of finding $\mathbf{p}$ aligned with, say, $z$, satisfies $\rho(\mathbf{r},\mathbf{p})=\rho(\mathbf{r},\mathbf{-p})$.
For this reason, the first non vanishing moment of $\rho$ is $\mathcal M_2=\int_{S^1}\mathbf{p}\otimes\mathbf{p}\rho$, which is taken to be the order parameter of nematic. We will define an order parameter for the triatic following the same logic, although there will be some differences. The tensor will be of rank 3, and the trace is not a well-defined operation when the rank is not 2. We will have to look for other invariants that signal the phase transition. In the following, $\rho$ is a positive-definite probability density on $S^1$. Other approaches that use a non-positive $\rho$ can be found in \cite{Virga:2015}.\\ 
At each point $\mathbf{r}$, triatic configurations are identified up to local rotations $R^k$ by $k(2\pi)/3$, and under reflections $H_k\equiv H(R^k\mathbf{p}), \ k=0,1,2$ across the legs of the triad. 
Correspondingly, the probability density has to satisfy four conditions: $\rho(\mathbf{r},\mathbf{p})=\rho(\mathbf{r},H_1\mathbf{p})=\rho(\mathbf{r},H_2\mathbf{p})$ and
\begin{align}
\label{eq:A1}
&\rho(\mathbf{r},\mathbf{p})=\rho(\mathbf{r},R\mathbf{p})=\rho(\mathbf{r},R^2\mathbf{p}).
\end{align}
Introducing an angular coordinate $\alpha$ on $S^1$, and writing $\mathbf{p}=(\cos\alpha,\sin\alpha)$, we find that~\eqref{eq:A1} can be rewritten as $\rho(\mathbf{r},\alpha)=\rho(\mathbf{r},\alpha+k\frac{2\pi}{3})$, $k=1,2$.
The first moment is $
\mathcal{M}_1(\mathbf{r})=\int_0^{2\pi}\ \rho(\mathbf{r},\alpha) \ (\cos\alpha,\sin\alpha)\, d\alpha 
$.
For later convenience we define the integration measure $
\mathtt{d}\mu\equiv  \rho(\mathbf{r},\alpha)d\alpha$.
As a consequence of~\eqref{eq:A1}, $\mathtt{d}\mu$ is invariant under shifts of $\alpha$ by multiples of $(2\pi)/3$.
We then split the integration interval in 3 sub-intervals from 0 to $(2\pi)/3$ to $(4\pi)/3$, so that $\mathcal{M}_1(\mathbf{r})=\vec{I}_0+\vec{I}_1+\vec{I}_2$, with
\begin{equation}
\vec{I}_j=\int_{j(2\pi)/3}^{(j+1)(2\pi)/3} d\mu (\cos\alpha,\sin\alpha) \quad j=0,1,2
\end{equation}
Now we make the change of variable $\alpha\to\alpha-j(2\pi)/3$. This doesn't affect the measure but brings the integration range to $[0,(2\pi)/3]$ and rotates $\mathbf{p}$ by $j(2\pi)/3$. The first moment is thus the sum of three vectors:
$\mathcal{M}_1(\mathbf{r})=\vec{I}_0+R\vec{I}_0+R^2\vec{I}_0=0$.
Notice that the symmetry under rotations alone was sufficient to constrain the first moment completely. 
A similar argument shows that the second moment $(\mathcal{M}_2)_{ij}=\int_{S^1}d\mu \, p_ip_j $ is a constant: $\mathcal{M}_2=\frac{1}{2}\texttt{Id}$.
Since the second moment is independent of the state of the system we go on to look at the third moment. This is not constant, and can be taken as the triatic's order parameter:
\begin{equation}
\label{eq:TT}
\mathbb{T}\equiv\mathcal{M}_3=\int_{S^1}d\mu \ \mathbf{p}\otimes\mathbf{p}\otimes\mathbf{p} \quad.
\end{equation}
For a completely disordered (or isotropic) phase, $\rho=\mathtt{constant}=(2\pi)^{-1}$. This gives 
$\mathbb{T}_\mathtt{iso}=0$.
On the other hand, the probability distribution of a perfectly ordered state cannot be $\rho=(2\pi)^{-1}\delta(\mathbf{p}-\mathbf{p_0})$, because it does not satisfy condition~\eqref{eq:A1}: $\delta(\mathbf{p}-\mathbf{p_0})\neq\delta(R\mathbf{p}-\mathbf{p_0})$ and $\delta(\mathbf{p}-\mathbf{p_0})\neq\delta(R^2\mathbf{p}-\mathbf{p_0})$.
This was not the case with a uniform nematic, because the Dirac delta function is even under inversion of its argument.
The analogy with nematics ends here, also because the trace is defined only on tensors of rank 2. 
A possible solution is to define the density in the isotropic phase as a sum of 3 delta functions:
\begin{equation}
\label{eq:3delta}
\rho_{ord}\sim\delta(\mathbf{p}-\mathbf{p_0})+\delta(\mathbf{p}-R\mathbf{p_0})+\delta(\mathbf{p}-R^2\mathbf{p_0})
\end{equation}
which is analogous to a distribution of three point masses if $\mathbf{p}$ was to represent a position in physical space. 
When we use \eqref{eq:3delta} in \eqref{eq:TT}, we find 
that $\mathbb{T}$ is automatically invariant under reflections across the three vectors of the triad. 
We will take the order parameter to be
\begin{align}
\nonumber
\mathbf{Q}=&\mathbf{p}\otimes \mathbf{p}\otimes \mathbf{p}\\
\label{eq:A4}
&+R\mathbf{p}\small{\otimes} R\mathbf{p}\otimes R\mathbf{p}+R^2\mathbf{p}\otimes R^2\mathbf{p}\otimes R^2\mathbf{p} \ .
\end{align}
 {The order parameter \eqref{eq:A4} describes accurately an ordered phase. Phase transitions and loss of orientational order are instead described by a scalar order parameter $S\equiv\braket{(\mathbf{l}\cdot\mathbf{n})\text{mod}(2\pi/3)}_\rho
\equiv\braket{(\cos3\alpha)}_\rho$, where $\mathbf{l}$ is the orientation of a single molecule, $\mathbf{n}=\braket{\mathbf{l}}_\rho$, with the average taken over a ball of radius $a\ll R\ll L$ ($a,L$ here are the molecular and system sizes respectively), and $\rho$ satisfies \eqref{eq:A1}. $S$ is obtained from the \emph{microscopic} $\mathbf{Q}$-tensor $\mathbf{l}\otimes\mathbf{l}\otimes\mathbf{l}$ by writing $\mathbf{l}=e^{i\alpha}$ and averaging over a ball: $S=\text{Re}\braket{e^{i3\alpha}}_\rho$.  A disorder-order phase transition would be described by a Landau free energy of the form $f(S)=\frac{1}{2}|\nabla S|^2+\frac{1}{2}a(T)S^2+BS^4$. Phase transitions, however, are not the focus of this paper. 
}
\section{Free Energy}
The ordered phase of the system is described by minimizers of a suitable free energy functional $F[\mathbf{Q},D\mathbf{Q}]=\int_\Sigma f(\mathbf Q,D\mathbf Q) d\mu(\Sigma)$, where the integration is taken over the space $\Sigma$ to which the liquid crystal is confined. The measure $d\mu(\Sigma)$ contains the determinant of the metric tensor on $\Sigma$. The free energy density is a scalar with respect to a change of coordinates, and thus it must be written in terms of invariants of the tensors $\mathbf{Q}$ and $D\mathbf{Q}$. The lowest order term $f_0$ containing $D\mathbf{Q}$ is 
\footnote{ {Notice that there are no cross-contractions between indices of $D$ and those of $Q$. Any other contraction between $D$ and $\mathbf{Q}$ vanishes because of the symmetries of $\mathbf{Q}$.}}:
\begin{equation}
\label{eq:A5}
f_0=\frac{1}{2}K \ D_aQ_{bcd}D^aQ^{bcd} \quad.
\end{equation}
 {For simplicity we choose to work in the one Frank constant approximation $K$, as the essential principles are revealed for $K_1=K_3$. $K$ is of the order $k_BT_m/a$, where $T_m$ is the melting temperature of the triatic and $a$ is the molecular size.} 
Consider flat $2-$dimensional space, choose polar coordinates $\{\mathbf{e}_1,\mathbf{e}_2\}\equiv\{\mathbf{e}_r,\mathbf{e}_\varphi\}$, and write $\mathbf{p}=\cos(\psi)\mathbf{e}_1+\sin(\psi)\mathbf{e}_2$.
There is only one independent component of the $\mathbf{Q}-$tensor~\eqref{eq:A4}: $\frac{3}{4} \cos(3\psi)=Q_{111} = - Q_{122}=- Q_{221} =- Q_{212}$, $\frac{3}{4}\sin(3\psi)=Q_{112} = Q_{121} =  Q_{211} = -Q_{222}$.
When expression~\eqref{eq:A5} is computed in flat polar coordinates $(r,\varphi)$, and assuming that $\psi$ depends on $\varphi$ only, the result is $f_0=\frac{81}{16}K \ r^{-2}(\p_\varphi\psi+1)^2$.
In terms of the angle $\theta$ between $\mathbf{p}$ and an horizontal cartesian axis,
\begin{equation}
f_0=\frac{81}{16}K\ r^{-2}(\p_\varphi\theta)^2 \quad.
\end{equation}
We will consider only the term $(\partial_\varphi\theta)^2$ because the prefactor contributes only to the core energy of the defect.
Can other invariants be included in the free energy? By computing the lowest order ones, we see that no invariant formed from contractions of $\mathbf{Q}$ alone has dynamical content. The reason lies in the symmetries of the order parameter, 
which forces any contraction to vanish: $
Q_{a\nu\nu}=Q_{a11}+Q_{a22}=0$ 
for any choice of $a=1,2$. Hence, terms like $Q^2$ or $Q^4$ vanish.\footnote{ This argument applies generally to any odd-rank tensor. For a 3rd rank tensor in three space dimensions, see \cite{Fel:1995}, Sec. IIIA.} Furthermore, contractions between $D$ and $\mathbf{Q}$ alone are not allowed, since $D$ and $\mathbf{Q}$ belong to different spaces. In view of this fact, we start our analysis of the ground state in flat space from the simple quadratic free energy functional
\begin{equation}
\label{eq:A8}
F=K\int d\mu (D\mathbf{Q})^2=K\int_0^\infty rdr\int_0^{2\pi}d\varphi(\partial_\varphi\psi+1)^2 \quad.
\end{equation}
Notice that, in terms of the unit vector $\mathbf{p}$, expression~\eqref{eq:A8} is equivalent to the functional $F=\frac{1}{2}\int_\Sigma d\mu(\Sigma)G_{ab}\partial_\mu p^a\partial^\mu p^b =\frac{1}{2}\int_{\Sigma} d\mu(\Sigma)(D\mathbf{p})^2$, 
that is usually encountered in nematic liquid crystals as well as in the non linear sigma model with a flat geometry  $(G_{ab}=\delta_{ab})$ in the field space. It can be proven that for any orientational order parameter of rank $p$, describing a $p-$fold symmetric LC, the free energy density $f_0$ always reduces to the nematic expression $(D\mathbf{p})^2$. We thus find that the free energy does not distinguish between different discrete symmetries of the order parameters, and the quadratic term $f_0$ is the same for all $p-$atic liquid crystals. This has an interesting consequence when we consider the class of odd-rank order parameters, including the triatic. These OP are odd under inversion of $\mathbf{p}$: $\mathbf{Q}(-\mathbf{p})=-\mathbf{Q}(\mathbf{p})$. In fact inversion seems to produce a distinct configuration of the LC. The free energy, however, is quadratic in $\mathbf{Q}$, and therefore shows an accidental symmetry under the same operation, suggesting that states related by inversion have at least degenerate energies. 
We will see in the following discussion that inversion in the OP space is equivalent to a proper rotation in physical $2-$dimensional space, so there is no degeneracy for the ground states, and both $\mathbf{Q}$ and $-\mathbf{Q}$ represent the same physical state.
\section{Ground states and defects}
The Euler-Lagrange equations for~\eqref{eq:A8} reduce to Laplace's equation in one dimension:
\begin{equation}
\label{eq:A9}
\partial_\varphi\partial_\varphi\psi(\varphi)=0
\end{equation}
The solutions in terms of the angle $\psi$ between $\mathbf{p}$ and $\mathbf{e}_r$ are linear functions of $\varphi$:
\begin{equation} 
\label{eq:A10}
\psi(\varphi)=a\varphi+\psi_0 \quad,
\end{equation} 
where $a\in\mathbb{R}$ and $\psi_0$ gives the initial orientation of $\mathbf{p}$ at $\varphi=0$. We can always choose coordinates in space such that $\psi_0=0$, so that $\mathbf{p}$ is parallel to the line $\varphi=0$.
\par We can classify solutions to~\eqref{eq:A9} according to their winding number, or topological charge, $s$, which measures the number of full revolutions of the director $\mathbf{p}$ along a closed path $\Gamma$ encircling the location of the defect. To do so, we must measure the angle $\varphi$ and the revolution angle of $\mathbf{p}$ with respect to the same reference axis. We thus introduce the angle $\theta(\varphi)$, related to $\psi(\varphi)$ through\footnote{The analysis of the defect structure follows the treatment of Landau and Lifshitz\cite{Landau:1986}} $\theta(\varphi)=\psi(\varphi)+\varphi $.
The winding number is then defined as $\oint_\Gamma d\theta=(2\pi)s$.
For $s=0$ the solution is free of defects, and describes a uniform distribution of triads: $\theta(\varphi)=\mathtt{const}$. The integral curves of the director are straight lines, which need to be identified with their images under $(2\pi)/3$ rotations  {(see Fig.1d).}
Notice that, unlike nematics, where the integral lines of $\mathbf{p}$ don't have an orientation, here each line carries a direction, dictated by the associated member of the triad. Orientation of the lines lifts the accidental $\mathbf{p}\to-\mathbf{p}$ symmetry of $F$, as was anticipated, because reversing orientations would require a rotation by $\pi/3$ in the OP space, which is not in the symmetry group of $\mathbf{Q}$.
\begin{figure}
\includegraphics[scale=.6]{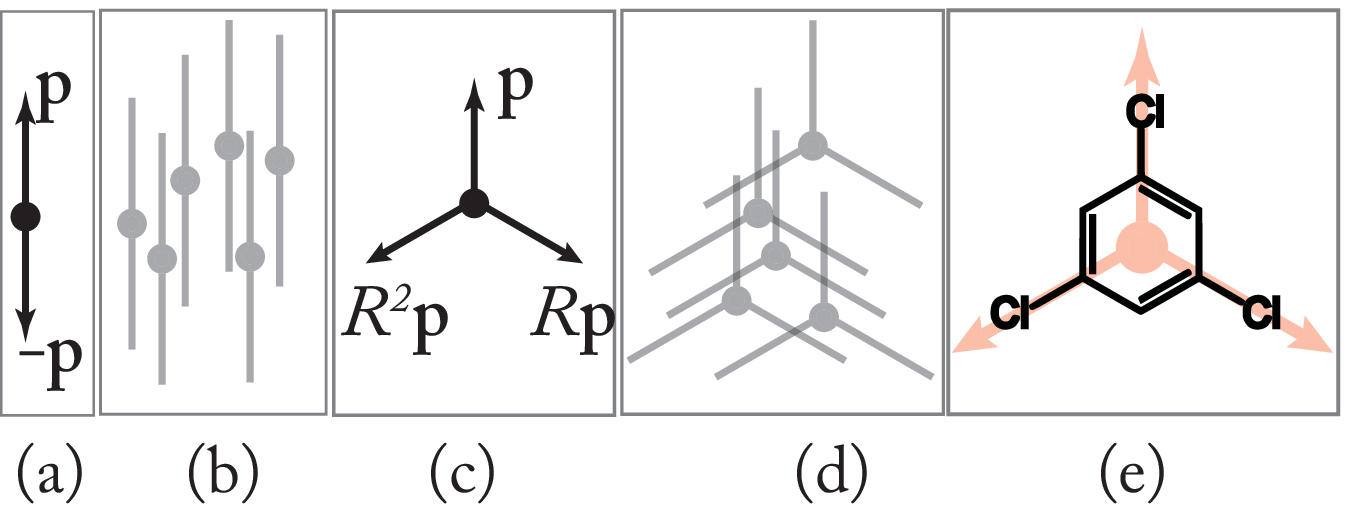}\\
 {{\bf Fig. 1} $-$ {\small {\bf{(a)}} Nematic frame, consisting of the vector $\mathbf{p}$ and its image rotated by $\pi$. {\bf{(b)}} Perfectly ordered nematic state in the plane. {\bf{(c)}} Triatic frame, consisting of $\mathbf{p}$ and its images rotated by $(2\pi)/3$ and by $(4\pi)/3$. {\bf{(d)}} Perfectly ordered triatic state in the plane.{\bf{(e)}} A possible realization of a triatic LC with 1,3,5-Trichlorobenzene. }}
\end{figure}
As was noticed before, there are two choices for the orientation of the integral lines but they are 
in fact the same configuration, because they can be made to coincide by rotating the physical space by $\pi/3$. So it is not surprising that the free energy associated with opposite orientations is the same. \\
The lowest, or elementary, topological charge allowed by the $3-$fold symmetry of triatics is $s=1/3$, because when $\varphi$ is rotated by $2\pi$ around a closed loop, the angle $\theta$ must come back to itself modulo an integer multiple of $1/3\cdot(2\pi)$:
$
\theta(\varphi+2\pi)=\theta(\varphi)+s(2\pi) \quad, s=\frac{1}{3}m$, $m\in\mathbb{Z} $
When this uniqueness condition is rewritten in terms of $\psi$, we find that the coefficient $a$ of equation~\eqref{eq:A10} is related to the charge by $a=s-1$.
The smallest charge is $s=+1/3$, and the corresponding defected state is $\psi(\varphi)=-\frac{2}{3}\varphi$.
For future convenience, we write explicitly the solutions $\psi$ and $\theta$ for an $s=1/3$ defect: $\theta(\varphi)=s\varphi=\frac{1}{3}\varphi$ and
\begin{align}
\label{eq:A11}
\psi(\varphi)&=(s-1)\varphi=-\frac{2}{3}\varphi=\frac{s-1}{s}\theta(\varphi)=-2\theta(\varphi) \quad.
\end{align}
The integral lines of $\mathbf{p}(r,\varphi)$ are found by requiring that $\mathbf{p}$ is parallel to the line element $d\mathbf{l}$. In polar coordinates, this reduces to
\begin{equation}
\label{eq:A33}
\frac{dl_\varphi}{dl_r}\equiv\frac{rd\varphi}{dr}=\frac{\sin\psi}{\cos\psi}=\tan\psi(\varphi) \quad,
\end{equation}
together with the condition that $\psi$ has the correct periodicity 
$\psi(\varphi+2\pi)=\psi(\varphi)+(s-1)(2\pi) $. 
Substituting~\eqref{eq:A11} into~\eqref{eq:A33}, we find
\begin{equation}
\label{eq:A12}
\frac{d\varphi}{d\log r}=\tan\biggl(-\frac{2}{3}\varphi\biggl) \quad.
\end{equation}
We can integrate~\eqref{eq:A12} by separating variables. We have to be careful about the sign of $\tan\psi$ as we complete one full revolution around the defect core. From~\eqref{eq:A11}, we find that as $\varphi$ changes from $0$ to $(2\pi)/3$, $\psi$ changes from $-(4\pi)/3$ to $0$. Within this interval, the function $\tan\psi$ changes sign 3 times. In particular $i)$ $\tan\psi<0$ for $\psi\in[-(4\pi)/3,-\pi]\cup[\pi/2,0]$ while $ii)$ $\tan\psi>0$ for $\psi\in[-\pi,-\frac{\pi}{2}[$.
As $r$ increases, the integral lines bend clockwise in case $i)$, and counterclockwise in case $ii)$. The function $\tan\psi$ can change sign in two ways. It can cross $\tan\psi=0$ at $\psi=-\pi$ and at $\psi=0$, in which case $d\varphi=0$. This means that the integral curve is a straight line, which will be called an \emph{Asymptote}. Or, it can jump from $+\infty$ to $-\infty$ at $\psi=-\pi/2$. This is a regular point for the differential equation, because $d\varphi/\tan\psi\to0$, so $r$ tends to a constant finite value. 
\par The integration of~\eqref{eq:A12} can be organized efficiently once we know the directions in the plane $\varphi=\tilde{\varphi}_m$ at which the tangent changes sign. These are found by inverting~\eqref{eq:A11}: $\tilde{\varphi}_m\equiv\varphi(\psi_m)\equiv-\frac{3}{2}(-m\frac{\pi}{2})$, $m=0,1,2 $.
For completeness, we observe that as we close the path $\Gamma$, $\varphi\to2\pi$, and $\psi\to-(4\pi)/3$. Hence the region below $x>0$ is regular, and the slope of the integral lines on the $x-$axis is everywhere equal to $\theta=(2\pi)/3$, which is what we expect for a $1/3$ defect.
From the previous discussion we conclude that:
\begin{itemize}
 \setlength{\itemsep}{1pt}
    \setlength{\parskip}{0pt}
    \setlength{\parsep}{0pt}
\item[a)] ${\varphi}\to 0^+$ is an asymptote ($\tan\psi\to 0^-$) 
\item[b)] $\tilde{\varphi}_1=\frac{3\pi}{4}$ divides behavior $i)$ from behavior $ii)$, and the curves are regular here
\item[c)] $\tilde{\varphi}_2=\frac{3\pi}{2}$ is an asymptote
\item[d)] For ${\varphi}\to2\pi$ the integral curves are regular, and their slope approaches $(2\pi)/3$ independently of $r$. 
\end{itemize}
 {Let us identify 3 regions in the ($r,\varphi$) plane: region I for $0<\varphi<(3\pi)/4$, region II for $(3\pi)/4<\varphi<(3\pi)/2$ and region III for $(3\pi)/2<\varphi<2\pi$.}
Integration of~\eqref{eq:A12} is easiest in region II, as here $\tan\psi$ is positive, so we can integrate $d\psi/\tan\psi=d\log(\sin\psi)$. The integral line containing the base point $(r_0,\varphi_0)$ is given by:
\begin{align}
&r_I(\varphi)=r_0\biggl[\frac{\sin(-2/3\varphi)}{\sin(-2/3\varphi_0)}\biggl]^{-3/2}
\text{, } \varphi,\varphi_0
\in\biggl[\frac{3\pi}{4},\frac{3\pi}{2}\biggl]
\end{align}
In regions I and III the tangent is negative, so we use the following procedure to integrate equation~\eqref{eq:A12}. We take region I as example. According to~\eqref{eq:A12}, here the derivative of $\varphi$ with respect to $r$ is negative. We define an auxiliary variable $u(\varphi)$ such that the value $\tan u$ is positive and
$\frac{d\varphi}{d\log r}=\tan\psi(\varphi)\equiv-\tan u(\varphi)$.
According to~\eqref{eq:A33}, the variable $\psi$ changes as $\psi(\varphi)=-2/3\varphi$. Inside region I, $\varphi$ changes between $0$ and $(3\pi)/4$.
The function $\tan x$ is odd under inversion of $x$. Hence, we define $u$ as the symmetric of $\psi$ with respect to the origin ($u(\varphi)=+2/3\varphi$), so that $\tan u>0$.
In terms of the variable $u$, the differential equation~\eqref{eq:A12} becomes
\begin{equation}
-\frac{3}{2}\frac{du}{\tan u}=d\log r \quad,
\end{equation}
which integrates to $r_{II}(\varphi)=r_0\biggl[\frac{\sin(2/3\varphi)}{\sin(2/3\varphi_0)}\biggl]^{-3/2}$ for $\varphi,\varphi_0
\in[0,\frac{3\pi}{4}]$.
Analogous considerations allow to solve~\eqref{eq:A12} in region III as well. The solution is identical to
$r_{II}$, but the interval of definition for $\varphi,\varphi_0$ is now $[3\pi/2,2\pi]$. The pattern of the full solution is represented graphically in  {Fig. 2 (Left panel).}
\begin{figure}
\begin{center}
\includegraphics[scale=0.20]{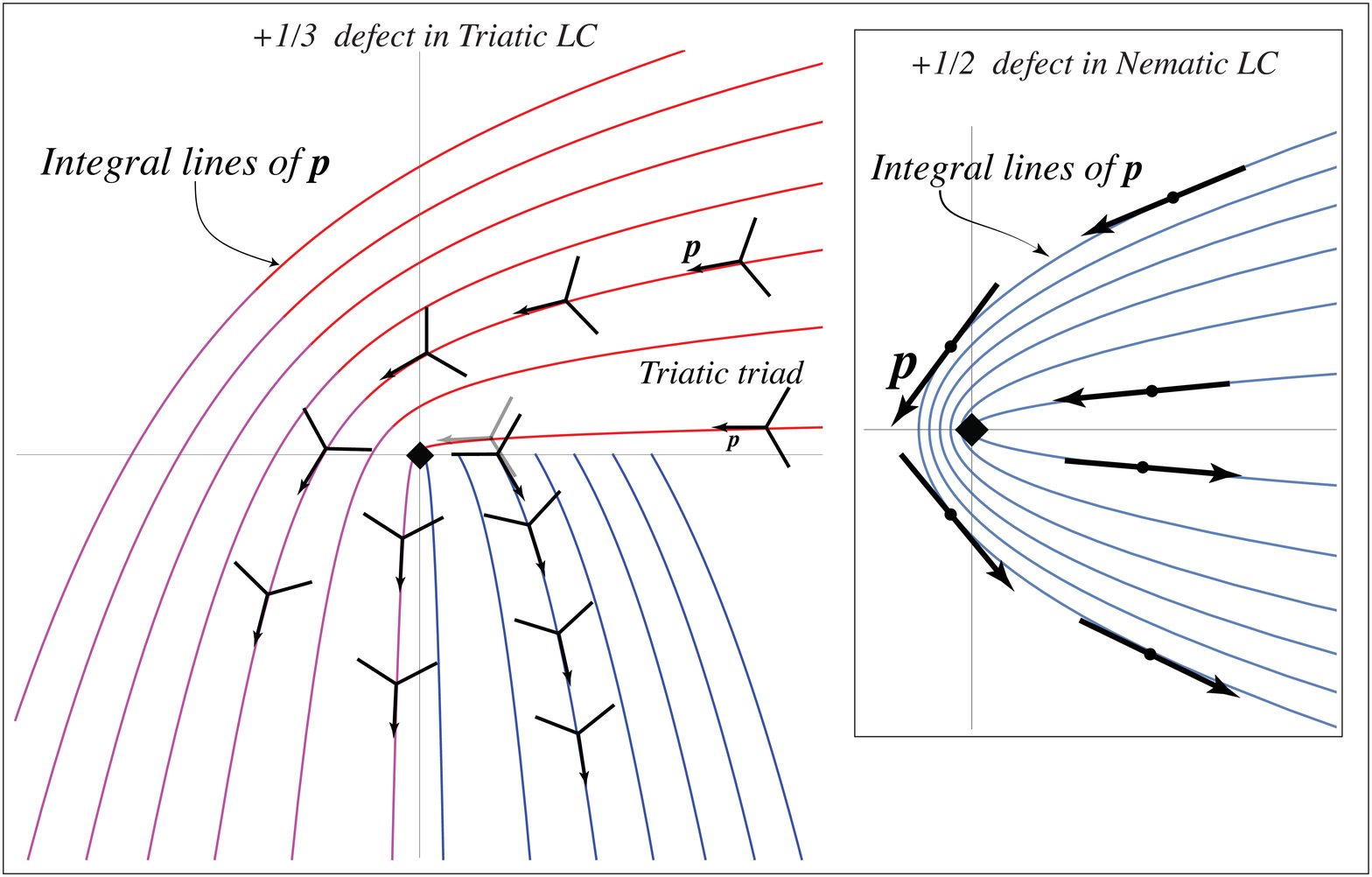}
\end{center}
 {{\bf{Fig.2}} $-$ $s=+1/3$ (Left panel): the defect structure of a triatic. (Right panel) The structure of a strength $+1/2$ nematic disclination. The vector $\mathbf{p}$ is used to solve \eqref{eq:A33} and then identified with the other two legs of the frame.}
\end{figure}
Each curve is the integral line of one leg $\mathbf{p}$ of the triad. Once the orientation of $\mathbf{p}$ is chosen along an integral curve, the orientation of the remaining two legs is uniquely determined. Imagine to draw $\mathbf{p}$ in region I, on the curve which is closest to the $x-$axis. We can choose to orient $\mathbf{p}$ to the right, so that $\theta=0$. As for the uniform, non-defected ground state, we observe again that each integral line carries an orientation,but an oriented pattern and its reflected image are connected by a rotation by $\Delta\varphi=\pi/2$ of the physical space, so there is no degeneracy, and no chirality associated with the pattern.
\par Now imagine a closed path $\Gamma$ that encircles the origin, and follow $\mathbf{p}$ along $\Gamma$. As $\Gamma$ crosses the negative $y$ axis, $\mathbf{p}$ has rotated by $\Delta\theta=\pi/2$. When the loop $\Gamma$ is approaching the starting point on the $x$-axis from below, the orientation of $\mathbf{p}$ approaches the limiting value $(2\pi)/3$. The reason is that the slope of the integral curves is precisely the angle $\theta$, see eq.~\eqref{eq:A33}. On the $x-$axis, the value of $\theta$ is $\theta(2\pi)=(2\pi)/3$, because the defect has charge $s=+1/3$. All the integral curves that approach the $x-$axis from below have slope $(2\pi)/3$ at $y=0$. 
 {The core profile $S(r)$ of the defect can be computed  in polar coordinates $(r,\varphi)$ using the scalar order parameter $S=\braket{\cos(3\alpha)}_\rho$  and imposing $S(r=0)=0,S(r\to\infty)=1$. The core energy scales with $2\log(L/a)$, so it is negligible compared to the bulk energy in the limit $L\to\infty$.}

 {In the following section we discuss triatic order in confined planar geometries, and make several predictions for the resulting equilibrium states of topological defects.  These could be tested experimentally.
 
\section{$p$-atics confined to a disc}

\begin{figure}
\begin{center}
\includegraphics[width=0.94\linewidth]{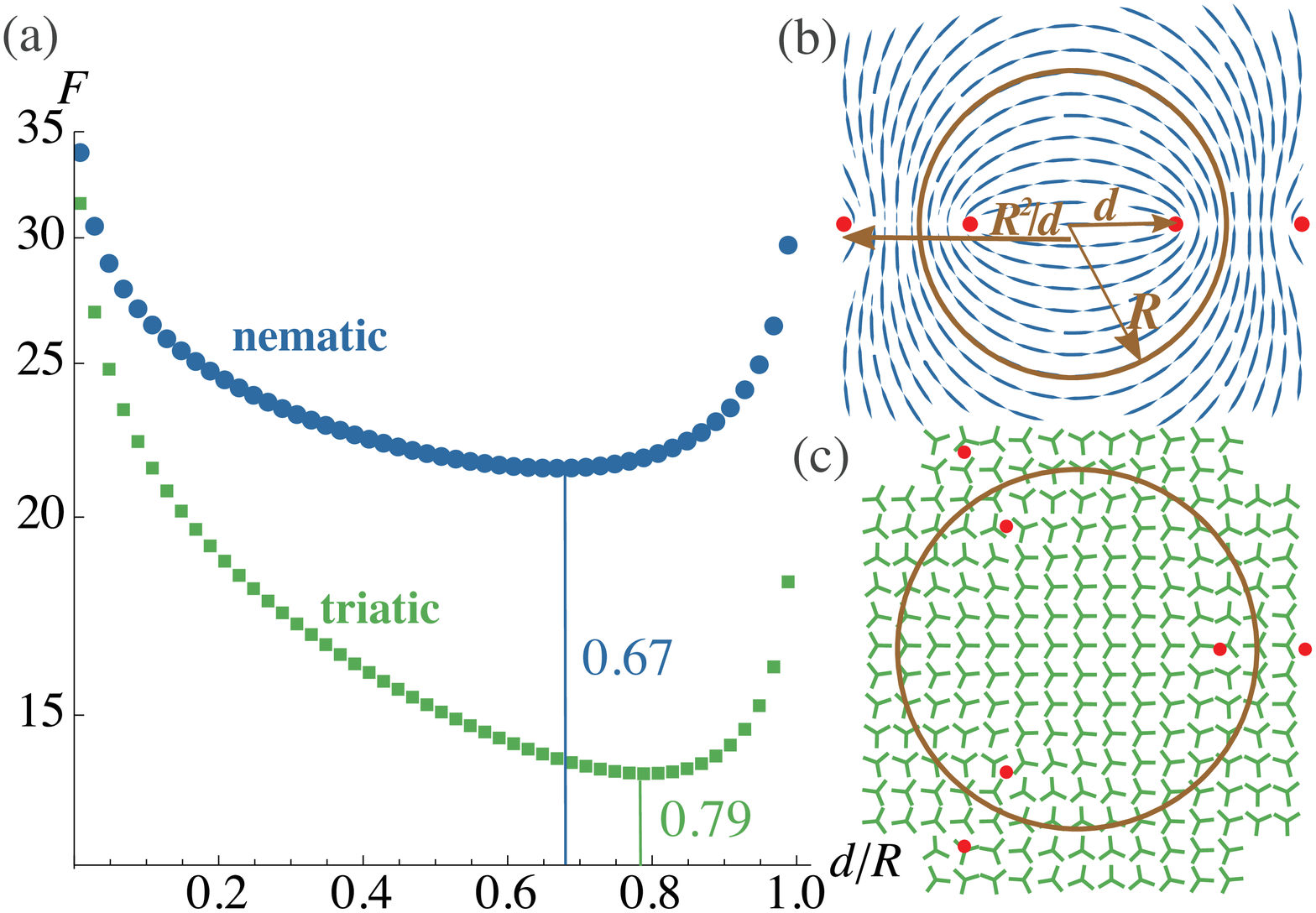}
\end{center}
{\label{fig:disc} {\bf{Fig.3}}$-$ \small{(a) The calculated free energy~\eqref{eq:A5} for nematic and triatic liquid crystals  confined to a disc. The tangential boundary conditions are achieved by placing $p=2,3$  charges of strength $+1/p$ at distances $d<R$ inside a disc and image charges at $R^2/d$ outside the disc of radius $R$~\eqref{eq:image}. The equilibrium  configurations for nematic (b) and triatic liquid crystals (c) occur for defects placed at $d/R=\{0.67,0.79\}$, respectively.}}
\end{figure}
 {
A $p-$atic LC confined to a disk of radius $R$ will develop $p$ defects of strength $1/p$ to screen the topological charge of the disk $\chi=1$. Thus, nematic and triatic LC will respectively form two $s=1/2$ and three $s=1/3$ defects that mutually repel each other. 
The location of these defects also depends on the boundary conditions. Strong anchoring of the prescribed orientation on the disk boundary is modeled by image charges $s'$ that push the bulk defects $s$ into the disk, away from the boundary. In equilibrium, the nematic is arrayed along a diameter at distance $d$ from the center (Fig.3b), while the three +1/3 defects of the triatic are positioned at the vertices of an equilateral triangle (see Fig.3c). The resulting equilibrium distance $d<R$ of the defects from the center of the disc depends on the precise number and configuration of interacting defects. The disk boundary is given by $z\cdot\bar z=R^2$ in complex coordinates $z,\bar z\equiv x \pm i y$. The orientation $\psi(z,\bar z)$ of $\mathbf p$, satisfies Laplace's equation $\p_{\bar z}\p_z\psi=0$ (see \eqref{eq:A5}) and the tangential alignment at the boundary $\psi|_{r=R}=\pi/2+\varphi$ is:
\be
\label{eq:image}
\psi(z,\bar z)=\arg\big[(z^p-d^p)^{1/p}\big(z^p-(R^2/d)^p\big)^{1/p}\big] .
\ee}
The argument, $\arg$, of the complex function is continuous  everywhere except at the points $d e^{i2\pi k/p}$ and $R^2/d e^{i2\pi k/p}$, with $k=0,1,2,\ldots, p-1$, located inside and outside the disc of radius $R$, where we place disclinations of charge $+1/p$ for a generic $p$-atic field. 
The free energy of the configuration~\eqref{eq:image} is minimized for $d/R=0.67$ for a nematic ($p=2$) and  $d/R=0.79$ for a triatic ($p=3$) - see Fig.3a.
The value of the integrated energy~\eqref{eq:A5} depends on the cut-off around the core of the defect (chosen to be $0.01 R$), but the location of the minimum depends only on the order $p$. This robust prediction was already verified experimentally for the nematic case, where elongated spindle cells~\cite{duclos:2016} organize in a structure analogous to Fig.3b with  the position of $+1/2$ disclinations  at $d/R=0.67$. Experiments with triatic LCs, for example using 1,3,5-Trichlorobenzene or Clathrin, would be highly desirable to verify the form of the equilibrium configuration and the structure of $+1/3$ disclination.
}
\section{Triatic LC confined to closed shells}
We have seen that the ground states of the free energy~\eqref{eq:A8} in flat space are ordered states that minimize bend and splay of their integral lines.  When a liquid crystal is confined on the $2-$dimensional surface of a vesicle, the substrate is also a dynamical variable, since it can adjust its shape to achieve the state of minimum energy. The energy cost to deform the surface is modeled through the Willmore functional $F_W$, that measures the energy required to increase the mean curvature $\vec{H}$ of the surface:
\begin{equation}
\label{eq:Willmore}
F_W[\mathbf{g}]=\frac{1}{2}\int_\Sigma \vec{H}^2 \, d\mu(\Sigma) \quad.
\end{equation}
The dynamical variable is the metric tensor on the surface $\mathbf{g}$, which is implicitly contained in the measure and in $\vec{H}$. The total free energy of a vesicle covered with a liquid crystal is therefore the sum of~\eqref{eq:A8} and~\eqref{eq:Willmore}  {\footnote{ {As with \eqref{eq:A5}, the one Frank constant approximation $K_1=K_3$ suffices to establish the existence of a polyhedral shell as the ground state.}}}:
\begin{equation}
\label{eq:ftot}
F[\mathbf{Q},\mathbf{g}]=\frac{1}{2}\int_\Sigma [K(D\mathbf{Q})^2 +\kappa \vec{H}^2 ] \, d\mu(\Sigma) \quad.
\end{equation}
 {The order parameter and the metric tensor are minimally coupled through the covariant derivative} $D_a=\partial_a+\mathbf{\Gamma}_a$, where  {$\mathbf{\Gamma}(\mathbf{g})$} are the Christoffel symbols on the surface, and depend on the metric tensor. The action of $D$ on the third rank tensor $\mathbf{Q}$ is $D_aQ^{bcd}=\partial_aQ^{bcd}+\Gamma^b_{ai}Q^{icd}+\Gamma^c_{ai}Q^{bid}+\Gamma^d_{ai}Q^{bci}$.
 {The core energy of defects is not included in \eqref{eq:ftot} because it is negligible for large system sizes. We imagine forming a vesicle with a triatic bilayer membrane immersed in a reservoir, so the surface tension is tuned to zero $(\delta F/\delta A=0)$ thanks to micro-transport (the area can freely fluctuate).}
According to expression~\eqref{eq:Willmore}, flat surfaces are favored by the Willmore energy, since $\Gamma^a_{bc}=0$ and $\vec{H}=0$ everywhere. A deformable closed vesicle with the topology of a 2-sphere, however, is subject to a topological constraint: the integrated Gaussian curvature must equal the Euler characteristic of the 2-sphere $\chi(\Sigma)=2$ Since $\mathbf{p}$ is taken to be a vector field in the tangent space of $\Sigma$, it must also contain a total topological charge equal to 2. Hence, the ground state of the order parameter on $\Sigma$ must necessarily contain defects whose sum is 2. Since the core energy of each defect is proportional to the square of the charge, $s^2$ (see \cite{chandra:1986}), the least energetic configuration contains 6 defects of charge $+1/3$. If we imagine to force the surface to be a sphere ($\kappa\to\infty$), and allow the defects to move on it, they will repel and try to maximize their mutual geodesic distance, occupying the 6 vertices of a regular octahedron (see \cite{lubensky:1992}). If the shape is now allowed to deform, the Willmore term will favor the formation of flat areas on the surface. In the limit $\kappa=0$, where the formation of sharp edges has no energy cost, we achieve a regular octahedron with $+1/3$ defects on each vertex. It is useful to check that there is no energy stored in the integral lines of $\mathbf{p}$ across the edges: as for Fig.1, the pattern is determined by the orientation of $\mathbf{p}$ with $\theta\in[0,(2\pi)/3]$. If we imagine to unfold the octahedron on a flat plane  {(Fig.4a)}, we will see that the lines are straight across the edges, hence they have zero bending energy into the tangent plane. 
 {We expect a variety of shapes for anisotropic bending rigidities.}
Thanks to the freedom to choose either orientation along the integral lines, we conclude that equation~\eqref{eq:ftot} allows two degenerate ground states. However, they have to be counted as the same object, because they are connected by a rigid proper rotation of the vesicle in $3-$dimensional space.
At each of the 6 vertices of the octahedron resides a $1/3$ defect in the triatic field, whose charge $s$ can be detected by travelling along a loop $\Gamma$ that encircles  vertex $V$, and measuring the net rotation of a reference leg in the triad, as explained in  {Fig.4b}. 
\begin{figure}
\ $\ \  $\ \  \includegraphics[scale=.32]{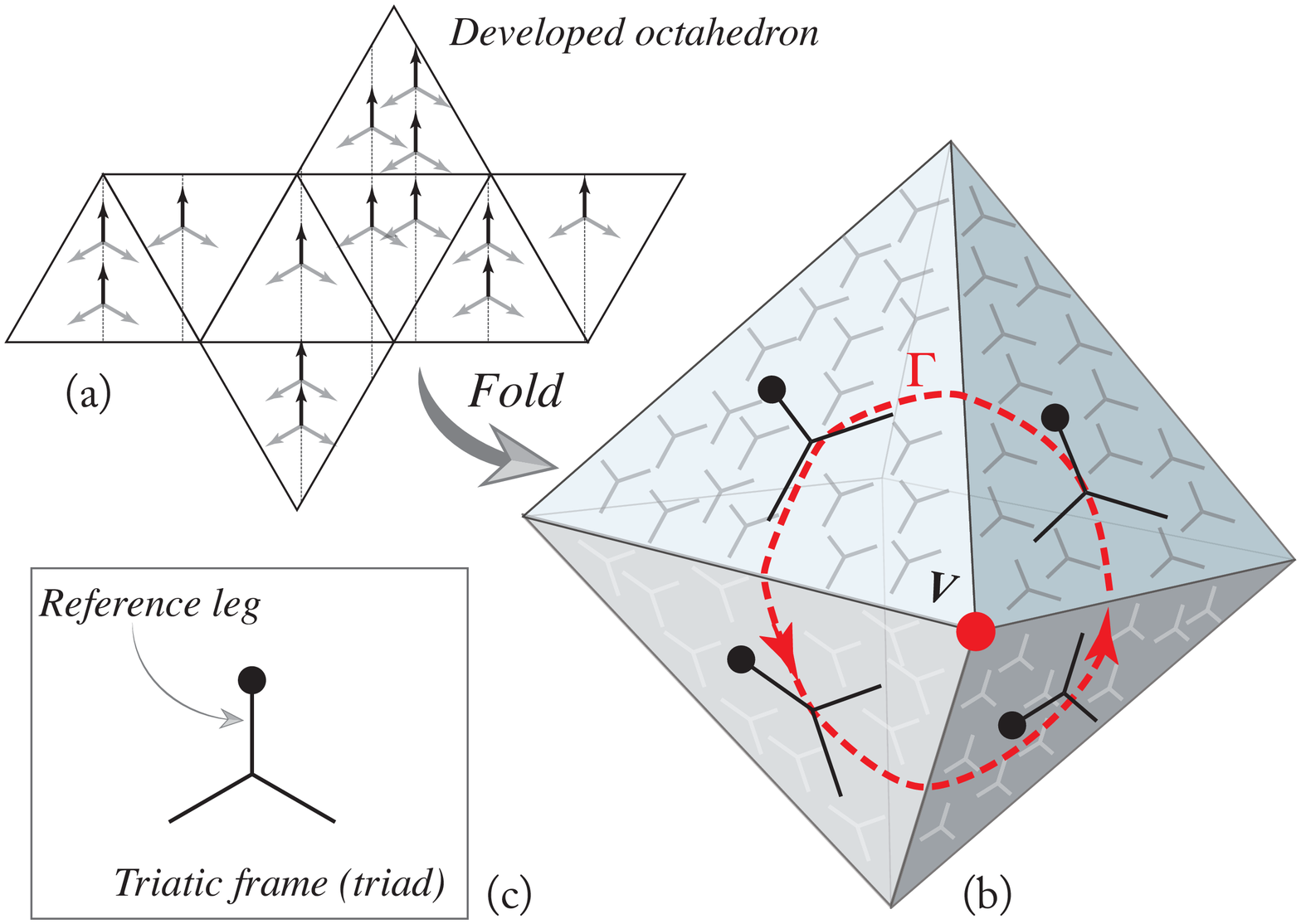}\\
 {
{\bf Fig. 4} $-$ {\small {\bf (a)} Unfolded octahedron with triatic order on its faces. The triatic LC is represented by a frame with 3 identical legs. The orientation of the frames is uniform within each face. {\bf (b)} Folded octahedron supporting triatic order. The triatic LC forms a defect of charge 1/3 around each vertex $V$ (red dot). Let $\Gamma$ be a closed path (red dotted line) encircling the vertex $V$. Choose a reference leg of the frame (indicated by a black dot in inset {\bf (c)} ), and follow its orientation as you travel around $\Gamma$. By the time we return to the starting point, we find the reference leg rotated by an angle of $(2\pi)/3$. }
}
\end{figure}
\section{Conclusions} We have constructed the order parameter describing a liquid crystal that breaks the isotropic group $O(2)$ down to $O(2)/\mathbb{Z}_3$ in the space of directions $\mathbf{p}$, finding that, unlike the nematic OP, it has vectorial properties. This distinction is a general feature of generalized liquid crystals. When the configurations of $\mathbf{p}$ are identified modulo $\mathbb{Z}_p$, the OP behaves like an element of the projective space for $p$ even, and as an ordinary vector when $p$ is odd. When the rank of the $\mathbf{Q}-$tensor is odd, we find that $\mathbf{Q}(-\mathbf{p})=-\mathbf{Q}(\mathbf{p})$. Using this property, we have shown that the functional expression of the free energy
reduces to the simple quadratic form $(D\mathbf{p})^2$, where $D$ is the covariant derivative. This expression is common to all $p-$atic LC.
For odd-symmetric LC, we have interpreted the accidental symmetry of the free energy under $\mathbf{p}\to-\mathbf{p}$ as the invariance of the system under proper rotations in the embedding space. 
In the specific case of triatics, the three-fold symmetry under rotations of $(2\pi)/3$ of the reference vector allows to construct an elementary defect of charge $+1/3$. We then considered a closed $2-$dimensional membrane coated with a triatic LC. We assumed that the total free energy takes into account both bending of the membrane, and elastic deformations of the vector $\mathbf{p}$ across the system. The state of minimum energy shows frustration between the constraint $\chi=2$ due to the spherical topology of the vesicle, and the tendency of $\mathbf{p}$ to be uniform across the surface. The energy is minimized by screening the total curvature charge with six elementary defects in the LC pattern ($2=6\cdot(1/3)$), whose locations maximize their mutual geodesic distance, as observed by Prost and Lubensky \cite{lubensky:1992} in 1992. In the limit of vanishing bending rigidity ($\kappa=0$), it is energetically favorable to form sharp straight edges between pairs of defects, and develop flat faces bounded by these edges  {through expulsion of gaussian curvature.} The resulting shape is an octahedron. The analysis employed to reach this conclusion, which was adopted earlier in \cite{Bowick:2012}, can be easily generalized to any $p-$fold symmetric liquid crystal. For example, in \cite{Man:2015} was constructed a non linear free energy for a vesicle covered with a four-fold symmetric LC, the tetratic (See also \cite{Li:2013},\cite{Geng:2009},\cite{Donev:2006}), which gives rise to a cubic shape. Hence, we expect that the ground states of generalized LC vesicles at zero bending rigidity realize all possible polyhedral shapes as one varies $p\in\mathbb{N}$. 
 {Precise predictions have been be made about the position of triatic LC defects confined to a disk. We hope that this work will encourage the design of experiments with biological and non-biological molecules of three-fold symmetric shape confined to two-dimensional layers to test such predictions. Clathrin or BTA molecules might be good candidates to test the unusual behavior of triatic LCs. Recent developments in DNA nanotechnology~\cite{Seeman:2002_1} have made possible the synthesis of molecules of various shapes, and triatic behaviour could be expected whenever an ensemble of such molecules exhibits local three-fold symmetry. Experimental tests of the effects of an anisotropic bending rigidity on the polyhedral shells would be highly beneficial to guide an extension of the present theory to less idealized situations. Emphasis should be put on the assembly of these polyhedral building blocks, through the functionalisation of the defects sites at the vertices.}

\acknowledgments
We acknowledge discussions with Suraj Shankar and Michael Moshe. This research was supported in part by the National Science Foundation under Grant No. NSF PHY-1125915. FS acknowledges the hospitality and stimulating discussions at the Kavli Institute for Theoretical Physics, University of California, Santa Barbara.

\bibliographystyle{eplbib}
\bibliography{ref}

\end{document}